# Superconductivity in the Topological Nodal-line Semimetal NaAlSi

Takahiro Yamada[1]*, Daigorou Hirai[2], Hisanori Yamane[1], and Zenji Hiroi[2]

[1] *Institute of Multidisciplinary Research for Advanced Materials, Tohoku University, Katahira, 2-1-1 Aoba-ku, Sendai 980-8577, Japan*
[2] *Institute for Solid State Physics, University of Tokyo, Kashiwanoha 5-1-5, Kashiwa, Chiba 277-8581, Japan*

NaAlSi is an *sp* electron superconductor crystallizing in a layered structure of the anti-PbFCl type with a relatively high transition temperature $T_c$ of ~7 K. Recent electronic state calculations revealed the presence of topological nodal lines in the semimetallic band structure, which attracted much attention owing to the superconductivity. However, experimental investigation remained limited because of the lack of single crystals. Here, we successfully prepared single crystals of NaAlSi by a Na–Ga flux method and characterized their superconducting and normal-state properties through electrical resistivity, magnetization, and heat capacity measurements. A sharp superconducting transition with a $T_c$ of 6.8 K is clearly observed, and heat capacity data suggest an anisotropic superconducting gap. Surprisingly, despite the *sp* electron system, the normal state is governed by the electron correlations, which is indicated by a $T^2$ resistivity and a Wilson ratio of 2.0. The origin of the electron correlation may be related to the orthogonal saddle-shaped Fermi surfaces derived from the Si $p_x$ and $p_y$ states, which intersect with the light Al *s* bands to form the nodal lines near the Fermi level. These results strongly suggest that the superconductivity of NaAlSi is not caused by a simple phonon mechanism but involves a certain unconventional aspect, although its relevance to the nodal lines is unclear.

## 1. Introduction

NaAlSi crystallizes in an anti-PbFCl-type layered structure with the centrosymmetric space group $P4/nmm$,[1]) in which Pb, F, and Cl atoms are replaced by Si, Al, and Na atoms, respectively, as depicted in Fig. 1: either set of the atoms forms a sheet with a square pattern and they stack along the *c*-axis. The Al and Si atoms are covalently bonded to each other to form a conducting layer made of the edge-sharing tetrahedra of Si atoms with the Al atom in the center. The Al–Si layers are separated by the double sheets of ionic Na atoms, giving rise to easy cleavage at the middle of the Na layers. The same structural type is found in the iron-based superconductors LiFeAs ($T_c$ = 16–18 K)[2,3] and LiFeP ($T_c$ = 6 K).[4] In 2007, Kuroiwa et al. found a superconducting transition at around 7 K via electrical resistivity and magnetization measurements on polycrystalline samples prepared under high pressure.[5] They also reported that an isostructural and isovalent NaAlGe did not show superconductivity above 1.8 K. Then, Schoop et al. showed that the $T_c$ of NaAlSi was increased up to ~9 K by applying a pressure of 2 GPa, decreased with further increase in pressure, and finally disappeared below 2 K above 4.8 GPa.[6] The reason for this suppression of superconductivity is unknown: their band structure calculations showed no significant change in the electronic structure under high pressure. Recently, Muechler et al. have measured magnetic penetration depth by muon spin rotation (μSR) and showed that its temperature dependence was explained by a conventional isotropic superconducting gap.[7]

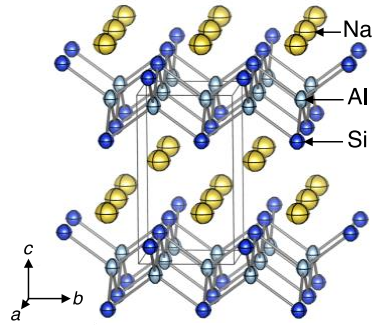

Fig. 1. (Color online) Tetragonal crystal structure of NaAlSi in the space group of *P*4/*nmm*. Our structural refinements using single-crystal X-ray diffraction data at 300 K yield lattice constants of *a* = 4.1217(1) Å and *c* = 7.3629(2) Å. On the basis of the refinements, each atom is drawn by a thermal displacement ellipsoid of 99% possibility. The atomic coordinates are [1/4, 1/4, 0.63461(8)] for Na, (3/4, 1/4, 0) for Al, and [1/4, 1/4, 0.20764(4)] for Si. Note that there are only two free atomic coordinates, the *z* parameters of Na and Al.

According to the first-principles electronic state calculations, NaAlSi is a naturally self-doped low-carrier-density semimetal consisting of the quasi-two-dimensional (2D) Al 3$s$ and Si 3$p$ bands reflecting the layered structure.[8] In 2019, three groups independently pointed out that NaAlSi is a topological nodal-line semimetal.[7,9,10] An overlapping between the electron-like Al $s$ bands and the hole-like Si $p$ bands produces multiple linear band crossings near the Fermi level, which results in a complex nodal-line structure protected by the crystalline symmetries of the space group, in the absence of spin–orbit coupling (SOC); the inclusion of SOC opens tiny gaps of less than 10 meV at all the crossing points, which can be neglected under finite thermal perturbation of experimental conditions.[10] Since the nodal lines appear slightly below the Fermi energy $\varepsilon_F$ and do not coexist with other fermions or extraneous bands, NaAlSi is a clean system realizing an excellent material platform for the investigation of the physics of the topological nodal-line semimetal.[9]

The mechanism of the superconductivity in NaAlSi has been discussed from a few viewpoints. The fact that NaAlGe is not a superconductor despite its similar electronic structure to NaAlSi suggests that the superconductivity is mediated by high-frequency phonons of Si with a lighter mass than Ge.[5,6] In fact, electronic and phonon calculations for NaAlSi gave an electron–phonon coupling constant $\lambda$ of 0.68 and a logarithmically averaged phonon frequency $\omega_{\ln}$ of 216 K, giving to a $T_c$ of 6.98 K.[11] On the other hand, a similarity to the lightly doped 2D nonmagnetic superconductors such as Li$_x$TiNCl,[12] in which the pairing mechanism may be ascribed to electron–electron interactions,[13] has been pointed out.[8] Moreover, a contribution of small electron pockets near the M point in the Brillouin zone to the superconductivity was inferred, because they were missing in NaAlGe.[8] However, the appearance of these electron pockets seems sensitive to the $z$ coordinates of Si, and they are missing also in NaAlSi in recent calculations.[7,9,10] Thus, the mechanism underlying the superconductivity of NaAlSi is still under debate.

The relationship between the superconductivity and nodal-line states of NaAlSi is intriguing to be focused on. Among several nodal-line semimetals thus far studied,[14] PbTaSe$_2$ is the only compound that has been extensively investigated in this context. It is a nodal-line Weyl semimetal with a noncentrosymmetric crystal structure and exhibits an $s$-wave superconductivity with $T_c$ = 3.72 K.[14] Topological surface states were predicted by first-principles electronic calculations and in fact observed by angle-resolved photoemission spectroscopy[15] and quasi-particle scattering interference imaging in scanning tunneling spectroscopy (STS).[16)] Moreover, a zero-bias conductance peak was observed at vortex cores in the superconducting state by STS, which was ascribed to a Majorana zero energy mode,[16)] predicted for topological superconductors.[17)]

The experimental information on the superconductivity and normal state of NaAlSi has remained limited, mainly because all the experiments were carried out on polycrystalline samples. In particular, there are no data on the anisotropy in normal and superconducting states. In addition, heat capacity data, which are often a key to understanding the superconducting characteristic, have not yet been obtained. In this study, we succeeded in growing single crystals of NaAlSi by a Na–Ga flux method and measured electrical resistivity, magnetic susceptibility, heat capacity, and Hall coefficient using these crystals. We show that the superconductivity and normal-state properties are not conventional and thereby intriguing physics in NaAlSi is waiting to be discovered.

## 2. Experimental Procedures

The Starting materials used for crystal growth were a Na lump (Nippon Soda 99.95%), Al rod (5 mm diameter; 99.9999%, Nilaco Co., Ltd.), Si powder (<150 mesh, 99.99%, Kojundo Chemical Laboratory Co., Ltd.), and Ga shot (99.9999%, Dowa Electronics). These materials were weighed at a molar ratio of Na:Al:Si:Ga = 3:1:1:0.5 and approximately 500 mg in total; the excess Na and Ga acted as a flux to produce a melt that could dissolve NaAlSi. The mixture was placed in a boron nitride crucible of 11 mm in diameter and 20 mm length (99.5%, Syowa Denko Co., Ltd.) and then enclosed in a stainless-steel container of 16.4 mm diameter and 80 mm length under Ar atmosphere in a glovebox. The container was heated in an electric furnace up to 1173 K in 3 h, cooled to 1073 K in 1 h, and cooled slowly to 823 K in 80 h, followed by subsequent furnace-cooling to room temperature without electrical power supply. Finally, the crucible was taken out from the SUS container in a glovebox, and the product was washed with alcohol to remove remaining Na. Many millimeter-sized platelet crystals of NaAlSi were collected from the bottom of the crucible together with a small amount of Na–Ga compounds. Dozens of NaAlSi crystals were obtained in one experimental run.

Quantitative elemental analysis of the single crystals was performed in an electron probe microanalyzer system (JEOL JXA-8200) with a wavelength-dispersive X-ray (WDX) spectrometer. To analyze the crystal structure at room temperature, X-ray diffraction (XRD) data were collected using Mo-Kα radiation ($\lambda$ = 0.71075 Å) and a single-crystal X-ray diffractometer (Bruker AXS, D8 QUEST). Data collection and unit cell refinement were carried out with the *APEX3* software package. Multiscan absorption correction to the observed data was carried out using the *SADABS* program,[18] and the crystal structure was refined using the *SHELXL*-2018/3 program installed in the *WinGX* software.[19,20] The crystal structures were visualized using the software package *VESTA*.[21]

Resistivity and Hall coefficient were measured in a physical property measurement system (PPMS) (Quantum Design Inc.). Indium metal was used to attain a good electrical contact with crystals. Heat capacity measurements were also carried out in a PPMS. Magnetization measurements were performed in a magnetic property measurement system (MPMS) (Quantum Design Inc.).

The electronic structure of NaAlSi was calculated on the basis of the density functional theory (DFT) using the "*Advance/PHASE*" software package (Advance Soft Corp., Japan). The lattice parameters and internal atomic coordinates determined by single-crystal XRD analysis in this study were adopted in the calculations.

## 3. Results

*3.1 Chemical characterizations and the crystal and electronic structures of NaAlSi*

*3.1.1 Chemical characterizations*

The grown single crystals of NaAlSi were bluish black in color and had a square plate shape with (001) facets of 1–4 mm width and 0.1–0.5 mm thickness [Fig. 2(a)]. It was found in previous studies that polycrystalline samples decomposed quickly in the ambient atmosphere,[1,5] while our single crystal showed better stability because of the small surface exposure. However, one has to be careful in handling the crystals because they violently react with water, sometimes igniting with white smoke. The crystal was stable in anhydrous ethanol and could be stored without degradation for more than a month. On the other hand, when it was put in concentrated hydrochloric acid for 3 h, it swelled along the *c*-axis direction and became an aggregate of thin plates in the shape of bellows [Fig. 2(b)], reflecting the two dimensionality in chemical bonding. This may be due to a gradual loss of Na atoms between the Al–Si layers following their reaction with hydrochloric acid or water.

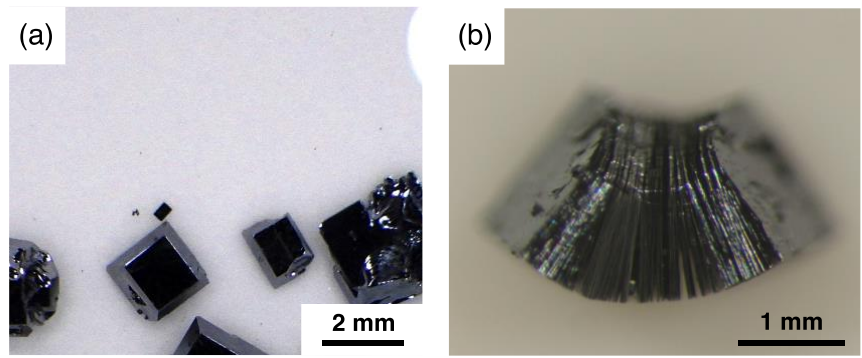

Fig. 2. (a) Single crystals of NaAlSi grown by the Na–Ga flux method. (b) Change in morphology of a crystal after being dipped in concentrated hydrochloric acid for 3 h.

The chemical composition of crystals was determined to be Na:Al:Si = 1.01(2):0.98(2):0.99(1) by the WDX spectroscopy, which is in good agreement with the stoichiometric composition. However, a trace of Ga of 0.48(6) at% (~1.4% per formula unit) was also detected, indicating that a small amount of Ga had been incorporated from the flux during the crystal growth. The Ga impurity may replace a portion of isovalent Al atoms. In fact, the Al content is slightly smaller than unity compared with the contents of Na and Si. Thus, carrier doping is not expected owing to the replacement.

*3.1.2 Crystal structure*

The crystal structure of NaAlSi was analyzed with a single crystal at 301(2) K on the basis of the previously reported crystal structure of tetragonal space group *P4/nmm* (No. 129).[1,6] The refinement converged with small reliability factors: $R1$ = 1.12% and $wR2$ = 3.19% (Appendix). The refined lattice constants are $a$ = 4.12170(10) Å and $c$ = 7.3629(2) Å, which are close to the values reported for a polycrystalline sample ($a$ = 4.119 Å, $c$ = 7.362 Å)[5] and a single crystal ($a$ = 4.1247(4) Å, $c$ = 7.3677(11) Å).[6] There are only two free parameters for the atomic coordinates: the $z$ values of Na and Si sites at the 2c and 2a Wyckoff positions, respectively. The refined $z$ coordinates are $z$(Na) = 0.63461(8) and $z$(Si) = 0.20764(4). These values are significantly different from the previous values of a single crystal analyzed by Westerhaus and Schuster [$z$(Na) = 0.622(6), $z$(Si) = 0.223(5)],[1] while they are close to the recent values of a single crystal analyzed by Schoop *et al*.

[$z$(Na) = 0.6346(3), $z$(Si) = 0.20759(15)][6] and calculated values via structural relaxation obtained by Tütüncü et al. [$z$(Na) = 0.6327, $z$(Si) = 0.2090].[11]

*3.1.3 Electronic structure*

The calculated band structure and the density of states (DOS) of NaAlSi are shown in Fig. 3, which are nearly identical to those previously reported.[7–10] Near the Fermi level, one dispersive conduction band of the Al 3$s$ character crosses with two valence bands of the Si 3$p$ character around the Γ point. These bands have small dispersions along the Γ–Z line, indicating a quasi-2D nature. The crossings between these bands form a complex nodal-line structure below the Fermi energy $\varepsilon_F$.[7,9,10] Here, we note flat valence bands appearing slightly above and below $\varepsilon_F$ at around the Γ and Z points, respectively. These bands have characters of Si $p_x$ and $p_y$: $p_x$ ($p_y$) forms a dispersive band whereas $p_y$ ($p_x$) forms a flat band along the Γ–X (Γ–Y) direction,[10] depending on the magnitudes of overlap integrals along the directions. As a result, a unique Fermi surface, which is a pair of orthogonal saddle-shaped Fermi surfaces extending along the X and Y directions, is generated with the fourfold rotation symmetry maintained, giving "fan-blade" hole surfaces;[8] note that if the equivalence of the two Fermi surfaces is lost owing to some reason, electronic nematicity could appear. These flat bands give a sharp peak in DOS at −0.12 eV below $\varepsilon_F$, as shown in Fig. 3(b). High thermopower was predicted because the slope of DOS at $\varepsilon_F$ is rather steep.[8] The DOS at the Fermi level $N(\varepsilon_F)$ is 0.90 states per eV and formula unit, which is slightly smaller than the previously calculated values of 1.0–1.2 states per eV and formula unit.[6,8,9,11] We will take the $N(\varepsilon_F)$ of 1.0 states per eV and formula unit as average and use for the following discussion.

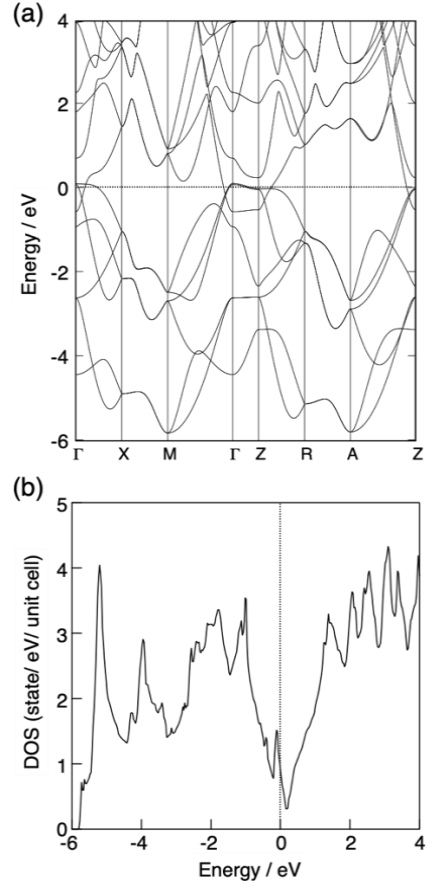

Fig. 3. (a) Band structure and (b) the DOS of NaAlSi calculated on basis of the structural parameters determined in this study.

*3.2 Superconducting properties*

*3.2.1 Meissner effect*

The temperature dependence of the magnetic susceptibility of a NaAlSi single crystal measured in a small magnetic field of 1 mT applied along the *a*-axis is shown in Fig. 4(a). A large diamagnetic response due to the Meissner effect is clearly observed below ~7.0 K. The shielding fraction determined by the zero field-cooling (ZFC) measurement at 2 K is approximately 120% of the perfect diamagnetism, and the Meissner fraction at 2 K in the field-cooling (FC) measurement is 8.7%. The former exceeds 100% owing to a demagnetization effect, and the latter is small because of the vortex pinning effect. Thus, the entire crystal must be rendered superconducting.

The isothermal magnetization initially increases linearly as shown in Fig. 4(b) and deviates at a field $H_{c1}^*$. The demagnetization factor $d$ was determined to be 0.294 by comparing the initial slope and $-1/(4\pi)$, and the lower critical field $H_{c1}(T)$ was obtained with $H_{c1}(T) = H_{c1}^{*}/(1-d)$. The temperature dependence of the thus-determined $H_{c1}(T)$ is shown in the inset of Fig. 4(b). A fit to the form $H_{c1}(0)[1-(T/T_c)^2]$ based on the Ginzburg–Landau (GL) theory yields a $H_{c1}(0)$ value of 4.7(1) mT, which is much smaller than the previous value of 10.1 mT reported for a polycrystalline sample.[5)] The GL magnetic penetration depth $\lambda_{GL}$, which is given by $[\Phi_0/2\pi H_{c1}(0)]^{1/2}$ with $\Phi_0$ being the magnetic flux quantum, is evaluated to be 260 nm.

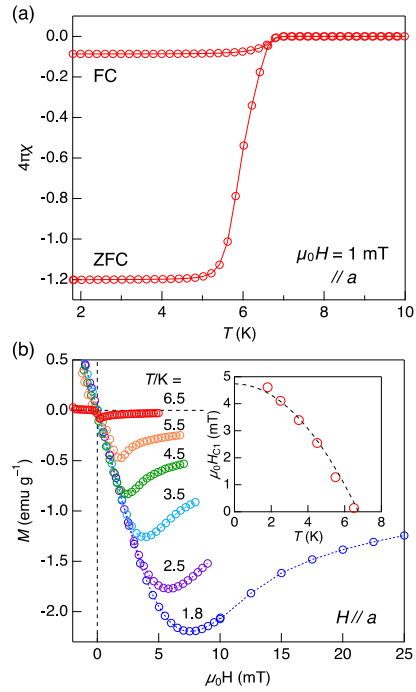

Fig. 4. (Color online) (a) Magnetic susceptibility of a NaAlSi single crystal measured upon heating, after cooling to 1.8 K in zero magnetic field (ZFC), and then upon cooling at a magnetic field of 1 mT applied along the $a$-axis (FC). (b) Isothermal magnetizations measured at 1.8–6.5 K with increasing magnetic field after ZFC. The inset shows the temperature dependence of the lower critical field $H_{c1}(T)$, in which the broken line represents a fit to the GL form $H_{c1}(T) = H_{c1}(0)[1-(T/T_c)^2]$, yielding $\mu_0 H_{c1}(0) = 4.7$ (1) mT.

*3.2.2 Superconducting transitions in resistivity*

The electrical resistivities along the $a$-axis ($\rho_a$) and $c$-axis ($\rho_c$) were measured on two different single crystals from the same batch and are shown in Fig. 5. The $\rho_a$ and $\rho_c$ values at 300 K are 1.66 and 19.7 mΩ cm, respectively, showing an anisotropy of a factor of 12, which is presumably attributed to the 2D nature of the electronic structure. Both the electrical resistivities decrease gradually upon cooling to 0.157 mΩ cm ($\rho_a$) and 1.99 mΩ cm ($\rho_c$) at 7.2 K, just above $T_c$. The residual resistivity ratios (RRRs) are both ~10, which are not large. We have examined five crystals prepared in three different runs and observed similar RRRs of 9–12. Although the scatter of RRR is small, the crystalline quality may still have room for improvement; one possibility is to reduce Ga contamination that may cause impurity scattering.

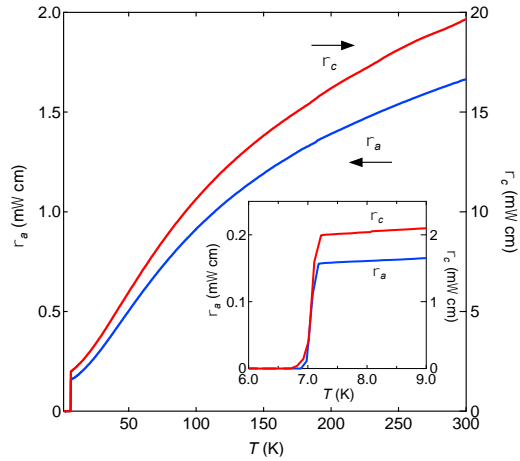

Fig. 5. (Color online) Electrical resistivities of NaAlSi single crystals with electrical currents parallel to the $a$-axis ($\rho_a$) and $c$-axis ($\rho_c$) at zero magnetic field. The inset shows superconducting transitions at ~7 K.

Both electrical resistivities decrease sharply from an onset of 7.2 K and reach zero below 6.8–6.9 K. The zero-resistance temperatures are close to the intrinsic $T_c$ value of 6.8 K determined from the heat capacity data as described later. To determine the upper critical field $H_{c2}$, magnetitic field dependences of isothermal resistivity with the electric current along the $b$-axis were measured. Under magnetic fields applied along the $a$-axis in the plane, the superconducting transition shifts gradually to the higher field side with decreasing temperature [Fig. 6(a)]. $H_{c2}(T)$ is determined at the midpoint of the transition and plotted in Fig. 6(c). Under magnetic fields along the $c$-axis perpendicular to the plane, two-step variations are observed at low temperatures and high magnetic fields [Fig. 6(b)]. The first sharp rise in resistivity with increasing magnetic field is due to a bulk superconducting transition as indicated by heat capacity measurements, while the second gradual recovery toward the normal-state value suggests that a part of the crystal remains in a low-resistive state in a temperature window even above the bulk $H_{c2}(T)$. It would be very interesting if this observation is related to surface superconductivity predicted for topological nodal-line superconductors.[22] We will return this issue with more detailed experiments in the near future.

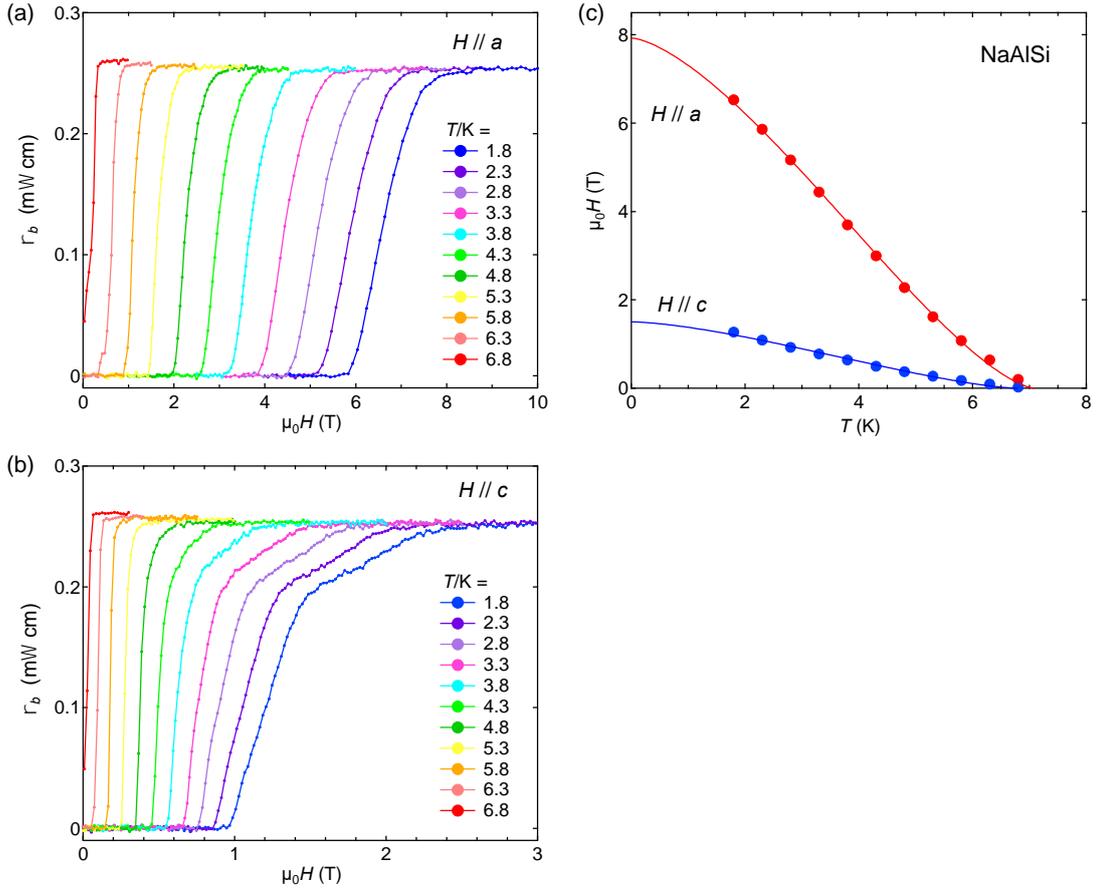

Fig. 6. (Color online) Isothermal electrical resistivities measured with an electric current along the $b$-axis as a function of magnetic field applied along the (a) $a$-axis and (b) $c$-axis. (c) Temperature dependences of the upper critical field $H_{c2}$ defined at the midpoints of the drops in resistivity.

$H_{c2}(T)$ is determined at the midpoint of the rise (the first rise for $H \parallel c$) and plotted in Fig. 6(c). Either of the thus-determined $H_{c2}(T)$ lines exhibits an upward curvature upon cooling from the zero-field transition temperature. Thus, the temperature dependences fit neither the GL form nor the Werthamer–Helfand–Hohenberg form applicable for most superconductors.[23] The upward curvature may be due to a Fermi surface effect for quasi-

2D superconductors.[24] Otherwise, it may be due to an intrinsically anomalous pairing mechanism. The previously proposed formula for a superconductivity with a local electron pairing, $H_{c2}(T) = H_{c2}(0)[1 − (T/T_c)^{3/2}]^{3/2}$,[25] has been successfully applied to many superconductors including PbTaSe$_2$[14] and borocarbides.[26] We used this formula to fit the whole $H_{c2}(T)$ data sets and obtained excellent fits with $\mu_0 H_{c2}(0) = 7.93(6)$ and $1.50(3)$ T for $H \parallel a$ and $c$, respectively. The former value is smaller than the Pauli limiting field of $1.84T_c = 12.4$ T, and the latter value is close to the reported $H_{c2}(0)$ value of 1.91T for a polycrystalline sample.[5] The anisotropy in $H_{c2}$ is approximately a factor of 5.3, reflecting the 2D electronic structure. This anisotropy is larger than that of the isostructural superconductor LiFeAs (ca. 1.5),[27] and is comparable to that of MgB$_2$ (ca. 5).[28] The anisotropic GL coherence lengths at zero temperature are determined on the basis of the GL theory to be $\xi_a(0) = 14.8$ nm and $\xi_c(0) = 2.8$ nm with an anisotropy ratio of 5.3: the anisotropic coherence length is given by $H_{c2} = \Phi_0/(2\pi\xi_a\xi_c)$ and $\Phi_0/(2\pi\xi_a^2)$ for $H \parallel a$ and $c$, respectively.[29] NaAlSi is a type II superconductor with a GL parameter $\kappa$ of $\lambda_{GL}/\xi_{GL} = 18$ for $H \parallel a$.

*3.2.3 Superconducting transition in heat capacity*

The heat capacity divided by temperature ($C/T$) of a NaAlSi single crystal is shown in Fig. 7(a). A sudden jump due to a second-order superconducting transition is observed at 6.5–7.0 K. The mean-field critical temperature determined by taking into account the entropy balance around the transition is 6.75 K, which coincides with the temperatures at which zero resistance occurs in Fig. 5. We measured the heat capacity of another crystal and obtained a $T_c$ of 6.80 K. Thus, the $T_c$ of NaAlSi is reliably determined to be 6.8 K.

The transition is completely suppressed below 2 K under an applied magnetic field of 1.5 T along the $c$-axis. The $C/T$ at 1.5 T is proportional to $T^2$ below ~8 K, as shown in the inset of Fig. 7(a), and can be fitted by the formula $C(T) = \gamma T + \beta T^3$, where $\gamma$ is the Sommerfeld coefficient and the second term is the Debye lattice contribution. A fitting in the range of 2.2–7 K yields $\gamma_{exp} = 2.166(8)$ mJ K$^{-2}$ mol$^{-1}$ and $\beta = 0.1022(3)$ mJ K$^{-4}$ mol$^{-1}$. The Debye temperature $\Theta_D$, which is given by $\Theta_D = (12\pi^4 NR/\beta)^{1/3}$ ($N$: the number of atoms per formula unit, $R$: the gas constant), is calculated to be 385 K, which is a reasonable value close to 320 K for a related compound CaAlSi.[30] The Sommerfeld coefficient $\gamma_{band}$ expected from the band structure is given by $\pi^2 k_B^2/[3N(\varepsilon_F)]$ and is calculated to be 1.18 mJ K$^{-2}$ mol$^{-1}$. Therefore, there is a moderate mass enhancement by a factor of 1.8.

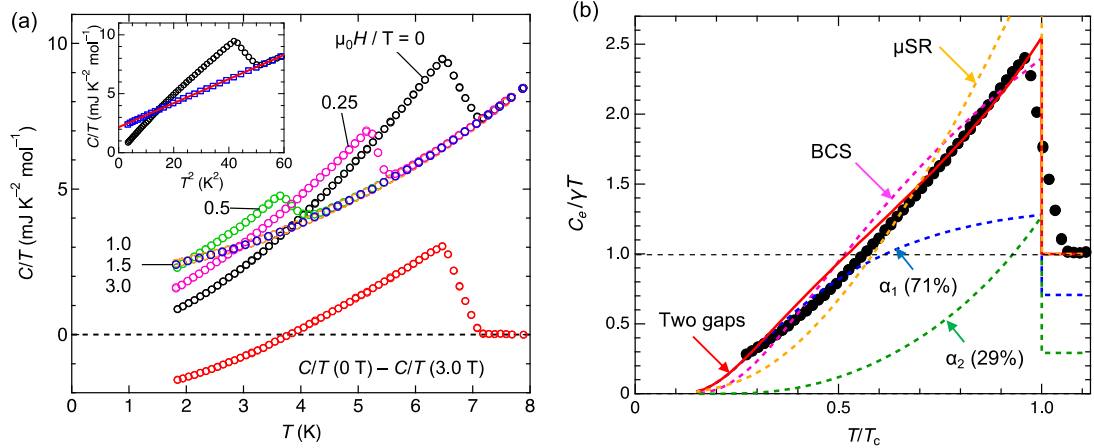

Fig. 7. (Color online) (a) Temperature dependences of heat capacity divided by temperature ($C/T$) in zero magnetic field and magnetic fields of 0.25, 0.5, 1.0, 1.5, and 3.0 T applied along the $c$ axis. A crystal of 11.25 mg weight was used. A difference between the 0 and 3.0 T data sets is also plotted at the bottom. The inset shows a $C/T$ versus $T^2$ plot, in which the solid line represents a linear fit of the 3.0 T data to the form $C(T) = \gamma T + \beta T^3$, yielding $\gamma$ = 2.166(8) mJ K$^{-2}$ mol$^{-1}$ and $\beta$ = 0.1022(3) mJ K$^{-4}$ mol$^{-1}$. (b) Electronic heat capacity $C_e$ divided by $T$ and $\gamma$ as a function of the reduced temperature $T/T_c$ with $T_c$ = 6.75 K. The red solid line represents a fit to the $\alpha$ model with two isotropic superconducting gaps, the separate components of which are shown by blue ($\alpha_1$ = 1.4, 71%) and green broken lines ($\alpha_2$ = 2.7, 29%). The magenta and orange broken lines represent $C_e/\gamma T$ for a BCS single gap with $\alpha$ = 1.765 and two gaps suggested by the previous µSR experiments ($\alpha_1$ = 1.0, 10%; $\alpha_2$ = 2.3, 90%),[7] respectively.

The temperature dependence of electronic heat capacity ($C_e/T$), which is obtained by subtracting the 3.0 T data as a reference for the lattice contribution and by adding $\gamma$, is shown in Fig. 7(b). The maximum value of jump in heat capacity ($\Delta C/\gamma T_c$) reaches 1.41, close to 1.43 expected for a weak-coupling BCS superconductor. However, $C_e/\gamma T$ significantly deviates from that expected for a weak-coupling BCS superconductor with a gap $\Delta_0/k_B T_c$ of 1.765 (magenta broken line). The measured values are smaller than the BCS curve in the intermediate temperature range, while they increase more rapidly with increasing temperature toward $T_c$, in contrast to the saturating tendency of the BCS curve. In addition, they tend to become larger than the BCS values toward $T$ = 0; the experimental value at 1.8 K is almost double of the BCS value. These results strongly suggest that the superconducting gap has larger and smaller magnitudes in part than the BCS value.

We have fitted the $C_e/\gamma T$ data with the $\alpha$ model that assumes two isotropic BCS-type superconducting gaps.[31)] The $\alpha$ model has been proved to reproduce heat capacity data appropriately for many superconductors such as $MgB_2$,[31)] $Nb_3Sn$,[32)] and $\beta$-pyrochlore oxides.[33)] It is assumed in the $\alpha$ model that two bands with different superconducting gaps, $\Delta_1$ and $\Delta_2$, contribute independently to the heat capacity. Each band is characterized by a partial Sommerfeld constant $\gamma_i$, so that the total $\gamma$ equals $\gamma_1 + \gamma_2$. Heat capacity data are fitted with three parameters: $\alpha_i = \Delta_i/k_B T_c$ ($i = 1, 2$) and $f = \gamma_1/\gamma$. Although the model assumes two gaps, it has been successfully applied to superconductors with a single anisotropic gap.[31)] By fitting the data of NaAlSi, we obtain $\alpha_1 = 1.4$ (contribution 71%) and $\alpha_2 = 2.7$ (29%) as one possible solution (red line), although the fitting has some ambiguity. The fit is good above $0.7T_c$, while not so in the lower temperature region. Although the fitting will be further improved by assuming three or more gaps, it may not be realistic, and one considers that there is a large directional anisotropy in the magnitude of the gap between 2.7 and nearly 0. On the other hand, two gaps have already been suggested from the temperature dependence of the magnetic field penetration depth determined by μSR experiments:[7)] $\alpha_1 = 1.0$ (contribution 10%) and $\alpha_2 = 2.3$ (90%), predominantly an $s$ wave with a large gap. However, the calculated heat capacity (orange broken line) does not agree with the experimental data at all. We guess that the μSR data were dominated by a part of the gap with a larger magnitude. Therefore, it is likely that NaAlSi has an anisotropic gap of the BCS magnitude on average, which is is deformed to have larger and smaller magnitudes depending on the direction or depending on the Fermi surfaces. This result means that the superconductivity of NaAlSi does not originate from a simple phonon mechanism.

### 3.3 Normal-state properties of NaAlSi
### 3.3.1 Temperature dependence of resistivity

Figure 8 shows the temperature dependences of in-plane electrical resistivity under a magnetic field of 4 T applied along the $c$-axis in the normal state. There is a small positive magnetoresistance, e.g., 7% at 7.5 K, which may be due to a normal magnetoresistance from the Lorentz force and thereby contributions from nodal lines are minimal. In fact, little magnetoresistance was observed for $H \parallel a$. We fitted the 4 T data below 15 K to the power law of $\rho^n$ and obtained $n = 2.07(1)$. A fitting to the equation of $\rho_0 + AT^2$ below 10 K yields $\rho_0 = 158.83(7)$ µΩ cm and $A = 0.201(2)$ µΩ cm K$^{-2}$ (red solid line in Fig. 8). Similar evaluations on a different crystal with RRR = 9 yielded $n = 2.11(3)$, $\rho_0 = 60.52$ (8) µΩ cm, and $A = 0.0897$ (5) µΩ cm K$^{-2}$. Therefore, NaAlSi exhibits a $T^2$ resistivity at low temperatures.

The $T^2$ resistivity has been known to occur for various systems with electron–phonon couplings that are extremely strong or electron correlations of any magnitude. For example, in A–15 compounds such as Nb$_3$Sn[34] and β-pyrochlore oxides such as KOs$_2$O$_6$,[35] the strong electron–phonon couplings cause the $T^2$ resistivity and strong-coupling superconductivity with a large jump in heat capacity at $T_c$. On the other hand, there are many correlated electron systems that exhibit $T^2$ resistivities at low temperatures.

We consider that the $T^2$ resistivity of NaAlSi originates from electron correlation effects, because the superconductivity is not in the strong-coupling regime (electron–phonon coupling must not be large). In fact, the observed $A$ values are close to those of strongly correlated electron systems such as V$_2$O$_3$ and uranium heavy-fermion compounds.[36] However, the $\gamma$ value of 2.17 mJ mol$^{-1}$ K$^{-2}$ for NaAlSi is significantly smaller than the large value of 32 mJ mol$^{-1}$ K$^{-2}$ for V$_2$O$_3$ or even larger values for heavy fermion compounds. As a result, the Kadowaki–Woods ratio $A/\gamma^2$ is obtained to be 0.04 or 0.015 µΩ cm (mol K/mJ)$^2$, much larger than the typical value of 10$^{-5}$ µΩ cm (mol K/mJ)$^2$.[37] This difference suggests that Coulomb interactions in NaAlSi are enhanced in a unique situation different from those of conventional correlated electron systems.

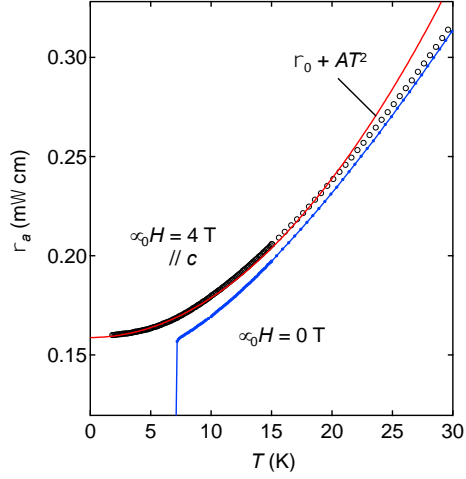

Fig. 8. (Color online) Electrical resistivity $\rho_a$ of a NaAlSi single crystal at low temperature below 30 K in magnetic fields of $\mu_0 H = 0$ and 4 T applied along the *c*-axis. The red solid line on the 4 T data represents a fit of the data below 10 K to the form $\rho_0 + AT^2$, which yields $\rho_0 = 158.83(7)$ μΩ cm and $A = 0.201(2)$ μΩ cm K$^{-2}$.

*3.3.2 Magnetic susceptibility*

Figure 9 shows a normal-state magnetic susceptibility for a NaAlSi single crystal in a magnetic field of 7 T applied along the *a*-axis ($\chi_a$) and *c*-axis ($\chi_c$) in the temperature range of 2–300 K. A large anisotropy in magnitude is observed: $\chi_a$ is 2.7 times larger than $\chi_c$ at 300 K. This anisotropy probably originates from the Landau orbital diamagnetism observed in 2D electron systems such as graphite.[38] The electrical resistivity of NaAlSi exhibits an anisotropy of a factor of approximately 10 (Fig. 5), and a small positive magnetoresistance is observed only in $\rho_a$ (Fig. 8). Thus, 2D electrons, probably light electrons from the Al *s* states forming a conducting layer, dominate the transport properties and must generate a sizable orbital diamagnetism that reduces only $\chi_c$. The magnitude of the orbital diamagnetism is estimated to be $-2.81\times10^{-5}$ cm$^3$ mol$^{-1}$ at 300 K from the difference between $\chi_a$ and $\chi_c$.

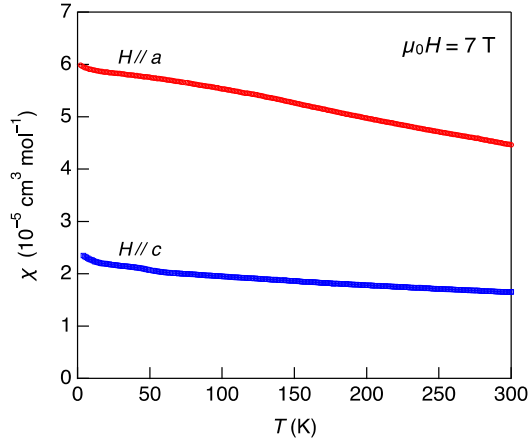

Fig. 9. (Color online) Magnetic susceptibilities $\chi$ of a NaAlSi single crystal measured in magnetic fields of 7 T applied along the $a$-and $c$-axes.

Both magnetic susceptibilities show considerable temperature dependences: they increase with decreasing temperature to $\chi_a = 6.0 \times 10^{-5}$ cm$^3$ mol$^{-1}$ and $\chi_c = 2.4 \times 10^{-5}$ cm$^3$ mol$^{-1}$ at 2 K. These temperature dependences must be due to the presence of a sharp peak in DOS just below $\varepsilon_F$, as shown in Fig. 3(b) by our electronic state calculation and also shown by previous studies.[6,8,9,11] In such a case, the DOS and thus the magnetic susceptibility should decrease with increasing temperature, because the number of thermally excited electrons with smaller DOSs than that at $\varepsilon_F$ increases with increasing temperature. Since the peak in DOS is located at 0.12 eV below $\varepsilon_F$, the reduction should occur even above room temperature.

Assuming that the contribution of the orbital diamagnetism is negligible for $\chi_a$ at the lowest temperature, the Pauli paramagnetism of NaAlSi is estimated to be $\chi_P = 6.1 \times 10^{-5}$ cm$^3$ mol$^{-1}$, after the subtraction of a small core diamagnetism of $8.5 \times 10^{-6}$ cm$^3$ mol$^{-1}$. On the other hand, the Pauli paramagnetism calculated from the DOS is $\mu_B^2 N(\varepsilon_F) = 1.62 \times 10^{-5}$ cm$^3$ mol$^{-1}$. Thus, the experimental $\chi_P$ value is 3.7 times larger than the calculated one. Moreover, band structure calculations gave a small value of exchange enhancement of the susceptibility by a factor of 1.17.[8] Hence, there is a relatively large enhancement in electron correlation that has not been considered in band structure calculations.

*3.3.3 Hall and Seebeck coefficients*

Figure 10 shows the field dependences of Hall resistivity $\rho_{xy}$ and the temperature dependences of Hall coefficient $R_H$ and carrier density $n$. $\rho_{xy}$ is proportional to the magnetic field with a negative slope at each temperature but, at low temperatures of 10 and 25 K, it slightly deviates above ~5 T. Thus, electrons dominate the transport while the contribution of holes becomes visible at low temperatures and high fields, which is consistent with the light electron bands and relatively heavy hole bands in the semimetallic band structure of NaAlSi.[8] The $R_H$ determined from the initial slope exhibits a significant temperature dependence, as expected for a semimetal, and reaches $9.8 \times 10^{-3}$ cm$^3$ C$^{-1}$ at 10 K. The electron carrier density estimated with the hole contribution being ignored is $6.4 \times 10^{20}$ cm$^{-3}$ at 10 K, which corresponds to 0.40% per formula unit. This value is much larger than those of semimetallic Bi (~$10^{17}$ cm$^{-3}$) or Dirac and Weyl semimetals and significantly smaller than those of conventional metals (~$10^{22}$ cm$^{-3}$).

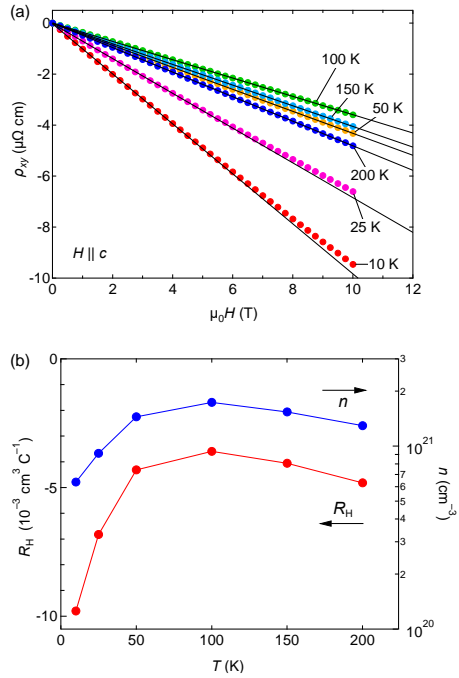

Fig. 10. (Color online) (a) Magnetic field dependences of Hall resistivity $\rho_{xy}$ with $H \parallel c$ and electrical current along the $a$-axis. (b) Temperature dependences of Hall coefficient $R_H$ and carrier density $n$.

We have also measured the Seebeck coefficient at room temperature using a polycrystalline sample and obtained a positive value of 32.5 µV K$^{-1}$. Thus, the Seebeck coefficient is dominated by the heavy holes in contrast to the Hall coefficient that is dominated by the light electrons. Since there is a sharp peak in DOS originating from the flat hole bands, one expects a large Seebeck coefficient.[8] However, the observed value is not so large probably because the peak in DOS is not so large with a small slope at $\varepsilon_F$.

## 4. Discussion

### 4.1 Mechanism of superconductivity in NaAlSi

We discuss the mechanism of superconductivity in NaAlSi by comparison with other related superconductors listed in Table 1. The $T_c$ of conventional superconductivity based on electron–phonon interactions is described by the Allen–Dynes modification of the McMillan formula: [39,40]

$$T_c = \frac{\omega_{\ln}}{1.2} \exp\left(-\frac{1.04(1+\lambda)}{\lambda - \mu^*(1+0.62\lambda)}\right), \qquad (1)$$

where $\omega_{\ln}$ is the logarithmically averaged phonon frequency, $\lambda$ is the electron–phonon coupling constant, and $\mu^*$ is the screened Coulomb pseudo-potential parameter. $\lambda$ is given by a product of the DOS at the Fermi surface $N(\varepsilon_F)$ and the pairing potential $V$. $N(\varepsilon_F)$ is proportional to the Sommerfeld coefficient $\gamma_{exp}$, and $\lambda$ is estimated from the relation $\gamma_{exp}/\gamma_{band} = (1+\lambda)$. $\omega_{\ln}$ can be estimated from the DOS of phonons, and $\mu^*$ is often assumed to be approximately 0.1 and causes ambiguity in the evaluation of $T_c$.

Table 1. Superconducting and normal-state properties of NaAlSi and selected superconductors.

| Materials | $T_c$ (K) | $\Delta C/\gamma T_c$ | $2\Delta/k_B T_c$ | $\gamma$ (mJ K$^{-2}$ mol$^{-1}$) | $\lambda$ | $\omega_{\ln}$ (K) | Comments | Ref. |
|---|---|---|---|---|---|---|---|---|
| NaAlSi | 6.8 | 1.41 | - | 2.17 | 0.8 | 145 | Anisotropic gap | This study |
| Al | 1.18 | 1.43 | 3.535 | 1.35 | 0.43 | 296 | Weak coupling | 41) |
| Pb | 7.20 | 2.71 | 3.95 | 2.98 | 1.55 | 56 | Strong coupling | 41) |
| MgB$_2$ | 38.7 | 1.32 | 4.0, 1.2 | 2.6 | 0.5 | 870 | Two gaps $\mu^* = 0.013$ | 31) |
| Li$_x$ZrNCl | 12.7 | 1.8 | 5.0 | 1.0 | 0.22 | 4900 | Low carrier | 12) |
| PbTaSe$_2$ | 3.72 | 1.41 | - | 6.9 | 0.74 | 112[*a] | Nodal-line SM | 14) |

[*a] Debye temperature.

Among the superconductors in Table 1, Al is a typical weak-coupling superconductor with a jump in heat capacity at $T_c$, $\Delta C/\gamma T_c$, of 1.43 and a superconducting gap $2\Delta/k_B T_c$ of 3.535, which are exactly expected from the BCS theory. Its $T_c$ is low at 1.18 K because of the small $\gamma$ and $\lambda$. Pb is a typical strong-coupling superconductor with its $T_c$ close to that of NaAlSi. However, it has a much larger jump in heat capacity at $T_c$ than NaAlSi and has a large isotropic gap of 3.95. The $\lambda$ of Pb is 1.55 (exceptionally large), which is due to low-energy phonons that can couple strongly with electrons. On the other hand, MgB$_2$ is a two-gap superconductor with gaps of 4.0 and 1.2 of nearly even weights as deduced from the α model.[31,42] Even if the $\lambda$ of MgB$_2$ is significantly smaller than those of Pb and NaAlSi, the $T_c$ is much higher at 38.7 K. When one attempts to explain this high $T_c$ using the above $T_c$ equation, $\omega_{\ln}$ becomes unrealistically large for $\mu^* = 0.1$ and 870 K for an unusually small value of $\mu^* = 0.013$.[31] This fact suggests that the equation is not applicable to MgB$_2$. Alternatively, it is considered that a strong electron–phonon coupling induced by a specific phonon, not an average of $\omega_{\ln}$, plays a critical role in determining $T_c$: the strong coupling of the B–B bond-stretching $E_{2g}$ branch of phonons to the bonding hole states of B 2$p$s may be responsible for the high $T_c$ in MgB$_2$.[42,43]

There is a reason to believe that the above-mentioned Al, Pb, and MgB$_2$ are phonon-based superconductors, whereas Li$_x$ZrNCl is clearly distinguished from them.[12] Although the heat capacity jump and the superconducting gap are large, both $\gamma$ and $\lambda$ are exceptionally small. To explain the high $T_c$, one has to assume a high-energy phonon of 4900 K. The characteristics of Li$_x$ZrNCl are the two dimensionality in electronic state and the low carrier density. Interestingly, $T_c$ increases with decreasing carrier density ($x$), followed by a sudden transition to an insulator below $x = 0.06$.[13] It is theoretically predicted that an exotic superconductivity based on plasmonic excitations occurs in the low-carrier limit in two dimensions, because the dynamical part of Coulomb interaction can change the repulsive Coulomb interactions to effective attractions.[44-46]

NaAlSi is a superconductor with an intermediate coupling constant of $\lambda = 0.8$. Provided $\mu^* = 0.1$, $\omega_{\ln}$ is to be 145 K to explain the $T_c$ in Eq. (1), which is almost two-thirds of the calculated value of 216 K from the phonon DOS;[11] $\lambda$ should be 0.68 for $\omega_{\ln} = 216$ K. Thus, a simple phonon mechanism may not work in NaAlSi. It is possible that a superconductivity based on a specific phonon occurs as in MgB$_2$. In fact, the $B_{1g}$ optical mode involving interlayer Si–Si vibrations along the $c$-axis, which has a high energy of 300 K, has a large $\lambda$ of 1.94.[11] On the other hand, a plasmon mechanism similar to that in Li$_x$ZrNCl has been suggested because NaAlSi is also a 2D conductor with low carrier density.[8] However, this may not be the case because the $\gamma$ and $\lambda$ values of NaAlSi are normal compared with the small values of Li$_x$ZrNCl.

Here, we point out the importance of the anisotropic superconducting gap and a possible role of electron correlations for the superconducting mechanism of NaAlSi. In general, because the electron–phonon interactions are always attractive, two electrons can come closer to each other, and the wavefunction of Cooper pairs has *s*-wave symmetry, as in the cases of Al, Pb, and MgB$_2$. In contrast, an anisotropic gap with nodes is preferred for pairing based on repulsive Coulomb interactions. The heat capacity data of NaAlSi suggest that the superconducting gap is likely anisotropic with $2\Delta/k_BT_c$ = 5.4 to ~0; the presence of nodes is not evident from our data. This anisotropy must originate from moderate electron correlations, as will be mentioned in the next section. Therefore, a certain exotic mechanism should work for realizing the superconductivity of NaAlSi.

*4.2 Electron correlations*

The Sommerfeld coefficient $\gamma_\text{exp}$ of 2.17 mJ K$^{-2}$ mol$^{-1}$ deduced from heat capacity measurements is larger by a factor of 1.8 than the value from band structure calculations. In contrast, the magnetic susceptibility is 3.7 times larger than the calculated one. Such imbalance in enhancements is characteristic for strongly correlated electron systems.[47] In fact, the Wilson ratio $R_W = (\pi^2/3)(k_B/\mu_0)^2(\chi/\gamma)$ is calculated to be 2.0, a typical value for them. Moreover, the observed $T^2$ dependence of resistivity is also a signature for Fermi liquids, although the Kadowaki–Woods ratio is unusually large. Hence, despite the *sp* electron systems, NaAlSi has an electronic state in which electron correlations are significantly enhanced.

One possibility is the electron correlations expected for nodal-line semimetals. For the Dirac or Weyl node, a long-range Coulomb interaction can be enhanced owing to less screening, and, for line nodes, an additional mass enhancement can occur along the line nodes.[48] In fact, such mass enhancements have been observed in the nodal-line semimetals ZrSiS[49] and ZrSiSe.[48] However, it is unclear whether such correlations are also effective in NaAlSi and cause the observed Fermi liquid behavior.

We consider that the origin is related to the saddle-shaped hole Fermi surfaces, the tops of which are located at 80 meV above $\varepsilon_F$ around the Γ point and just below $\varepsilon_F$ around the Z point.[8,10] The highly dispersive Al 3*s* electron bands intersect these Si 3*p* flat hole bands, generating Dirac crossing points forming complex nodal lines. These light electrons dominate the transport properties of NaAlSi, as indicated by the Hall coefficient. However, they must become heavier by dragging the heavy holes in the flat bands, resulting in a moderate electron correlation. The observed anisotropy in a superconducting gap suggests that this electron correlation plays an important role in Cooper pairing. A microscopic mechanism calls for future theoretical and experimental study. The role of nodal lines and surface states shold also be investigated in the future.

## 5. Conclusions

The superconducting and normal state properties of NaAlSi, which is a quasi-2D $sp$ semimetal with topological nodal lines, were investigated by electrical resistivity, magnetic susceptibility, heat capacity, and Hall coefficient measurements using single crystals grown by the Na–Ga flux method. A bulk superconductivity was observed below $T_c$ = 6.8 K with an anisotropy by a factor of 5.3 in the upper critical field. Heat capacity data reveal an anisotropic superconducting gap, suggesting an unconventional superconducting mechanism. The normal state possesses an anisotropy by a factor of 10 in resistivity and is characterized by an electron correlation, unusual for $sp$ electrons, which is evidenced by a Wilson ratio of 2.0 and $T^2$ resistivity. The origin of the electron correlation is suggested to be the flat saddle-shaped hole Fermi surfaces originating from Si $p_x$ and $p_y$ states. The superconducting mechanism of NaAlSi may be related to this electron correlation and is clearly distinguished from those of other materials owing to the unusual electronic setup.

**Acknowledgment**

This research was conducted with the support of the joint research in the Institute for Solid State Physics, University of Tokyo. It was financially supported by JSPS KAKENHI Grants (JP20H02820 and 20H05150). The authors would like to thank Yumi Oikawa, Reina Kusaka, and Chikako Nagahama for their help in the sample preparation and Takashi Kamaya (Tohoku University) for the EPMA measurements.

**Appendix A**: Crystal structure data and the results of refinement for NaAlSi.

| | |
|---|---|
| formula weight $M_r$ / g mol$^{-1}$ | 78.06 |
| crystal form, color | platelet, blueish black |
| crystal size / mm$^3$ | 0.047 × 0.095 × 0.102 |
| crystal system | tetragonal |
| space group, $Z$ | $P4/nmm$ (No. 129), 2 |
| radiation wavelength $\lambda$ / Å | 0.71073 (Mo-$K\alpha$) |
| $F_{000}$ | 76 |
| temperature $T$ / K | 301(2) |
| unit cell dimensions | |
| $a$ / Å | 4.12170(10) |
| $c$ / Å | 7.3629(2) |
| unit cell volume $V$ / Å$^3$ | 125.084(7) |
| calculated density $D_{cal}$ / Mg m$^{-3}$ | 2.073 |
| absorption coefficient $\mu$ / mm$^{-1}$ | 1.047 |
| limiting indices | |
| $h$ | $-6 \leq h \leq 6$ |
| $k$ | $-6 \leq k \leq 6$ |
| $l$ | $-12 \leq l \leq 12$ |
| $\theta$ range for data collection / ° | 2.77–36.29 |
| reflections collected, unique ones | 1918, 210 |
| $R_{int}$ | 0.0234 |
| data, restraints, parameters | 210, 0, 10 |
| weight parameters; $a$, $b$ | 0.0193, 0.0025 |
| extinction coefficient $x$ | 0.07(2) |
| goodness-of-fit on $F^2$; $S$ | 1.180 |
| $R1$, $wR2$ [$I > 2\sigma (I)$]* | 0.0107, 0.0317 |
| $R1$, $wR2$ (all data)* | 0.0112, 0.0319 |
| largest diff. peak and hole / e·Å$^{-3}$ | 0.152, −0.176 |

*$R1=\sum||F_o|-|F_c||/\sum|F_o|$, $wR2 = \{[\sum w[(F_o)^2 - (F_c)^2]^2] / [\sum w(F_o^2)^2]\}^{1/2}$; $w = [\sigma^2(F_o)^2 + (aP)^2 + bP]^{-1}$, where $P = [(F_o)^2 + 2(F_c)^2]/3$.

**Appendix B**: Atomic coordinates, the equivalent isotropic displacement parameter $U_{eq}$ and the anisotropic displacement parameters $U_{11}$ and $U_{33}$ in the unit of Å$^2$ for NaAlSi.

| Atom | Site | $x$ | $y$ | $z$ | $U_{eq}$ | $U_{11}$* | $U_{33}$ |
|---|---|---|---|---|---|---|---|
| Na | 2c | 1/4 | 1/4 | 0.63461(8) | 0.02075(13) | 0.02231(19) | 0.0176(2) |
| Al | 2a | 3/4 | 1/4 | 0 | 0.00995(10) | 0.00875(12) | 0.01237(14) |
| Si | 2c | 1/4 | 1/4 | 0.20764(4) | 0.01010(10) | 0.01026(11) | 0.00978(13) |

*$U_{11} = U_{22}$, $U_{23} = U_{13} = U_{12} = 0$

*E-mail: takahiro.yamada.b4@tohoku.ac.jp